\documentclass[12pt]{iopart}
\usepackage{graphicx}
\usepackage{epstopdf}
\usepackage{color}
\usepackage{cite}

\renewcommand{\figurename}{Figure}

\begin{document}

\title[Harmonic generation for parallel propagation of EM  wave  in  a magnetized plasma]
{Harmonic generation in magnetized plasma for  Electromagnetic  wave propagating parallel to  external magnetic field}

\author{Trishul Dhalia$^{*1}$, Rohit Juneja$^*$, Laxman Prasad Goswami$^*$, Srimanta Maity$^{**}$ and Amita Das$^{*2}$}

\address{$^{*}$Department of Physics, Indian Institute of Technology Delhi, Hauz Khas, New Delhi-110016, India \\}
\address{$^{**}$ Extreme Light Infrastructure ERIC, ELI Beamlines Facility, Za Radnicí 835, 25241 Dolní Břežany, Czech Republic}
\ead{$^{1}$trishuldhalia@gmail.com, $^2$amita@iitd.ac.in}
\vspace{10pt}

\begin{abstract}
 The harmonic generation has always been of fundamental interest in studying the nonlinear nature of any physical system. In the present study, Particle - In - Cell (PIC) simulations have been carried out to explore the harmonic generation of Electromagnetic waves in a magnetized plasma. The EM wave propagation is chosen to be parallel to the applied external magnetic field. The simulations show the excitation of odd higher harmonics of RCP (Right circularly polarized) and LCP (Left circularly polarized) when the incident wave is linearly polarised. The harmonic generation is maximum when the incident EM wave frequency matches the electron cyclotron frequency. When the incident EM wave has a circular polarization, no harmonics get excited. A theoretical understanding of these observations has also been provided. The studies thus show that by appropriately tailoring of plasma parameters EM waves of higher frequencies and desired nature of circular polarization can be generated. 
\end{abstract}

\section{Introduction}\label{sec:Introduction}
Electromagnetic wave (Lasers, Microwave, and Radio-frequency regime) is used in a variety of contexts. 
The importance and applicability of laser plasma interactions are well-known and strongly pursued in the context of fusion \cite{kaw2017nonlinear,das2020laser}, particle acceleration \cite{nishida1987high,joshi2006plasma}, etc. Microwaves are also being employed for many studies such as   the generation of plasma sources \cite{tarey2016studies,ganguli2016development,ganguli2019evaluation}. Its absorption leads to plasma heating \cite{litvak1993nonlinear}, which is important in several contexts such as tokamak \cite{gilgenbach1980heating,mueck2007demonstration} , stellarator \cite{koehn2012schemes,hammond2018overdense} , particle acceleration \cite{alvarez1981application,batanov1986large}, microwave heated chemical reactions \cite{tiwari2020microwave} and mirror machines \cite{launois1972contribution, ganguli1997characterization}. While high-power lasers have already been available for a long time,  the production of high-power microwave sources has been recent  \cite{levush1996high,fan2020direct,wang2020preliminary}.  In the recent study by \cite{xiao2020efficient,wang2020preliminary}, high power microwave pulsed source of up to 4.6 GW and of frequency 9.96 GHz has been achieved in  experiments. These high-power microwave sources have stimulated interest in the study of  nonlinear effects  in microwave plasma interaction.  Nonlinear phenomenon such as self-focusing and self-guiding \cite{ito1996formation,ito2004propagation}, resonant absorption \cite{lee1982hot,rajyaguru2001observation}, soliton excitation \cite{nishida1986excitation,kaw1992nonlinear} and ponderomotive forcing have been studied \cite{max1974self,max1976strong}  by several authors. 

The studies in laser-plasma interaction have primarily been carried out for the unmagnetized plasma response. The magnetic field requirement is considerably high (and hence not possible to achieve) for eliciting magnetized response at laser frequencies from the charged species of the plasma. However,  recent technological development of  producing  high magnetic fields of the order of kilo Tesla in the laboratory \cite{nakamura2018record} and proposals to produce even stronger, of the order of Mega Tesla \cite{korneev2015gigagauss} magnetic fields, has sparked research interest in the area of laser interacting with magnetized plasmas for which a variety of  theoretical and simulation studies are  now being conducted 
\cite{vashistha2020new,vashistha2021excitation,vashistha2022localized,mandal2020spontaneous,mandal2021electromagnetic,kumar2019excitation,goswami2021ponderomotive,goswami2022observations,maity2021harmonic,PhysRevE.105.055209}. On the other hand,  the  available microwave and RF sources typically have low power and produce continuous waves. The magnetized plasma response can be invoked relatively easily at low MW frequencies in experiments with reasonably smaller magnetic field strength values. The MW plasma experiments involve low-power microwave sources and complicated configurations of magnetic fields. Thus the conventional laser plasma and the MW/RF research, in general, have explored distinct regimes. The former (laser) has been extensively employed to study an unmagnetized and highly nonlinear response from the plasma to EM waves. On the other hand, MW/RF has typically explored magnetized but, at best, a weakly nonlinear response. Also, the explorations of MW/RF have primarily been conducted in the context of Tokamak devices and have, therefore, employed complex magnetic field geometry. This gap in the regime of research explorations between laser and MW/RF is now getting bridged by technological advancements in the production of strong magnetic fields and pulsed high-power MW sources.

  
One of the nonlinear effects associated with laser interaction with plasmas is the generation of higher harmonics. Excitation of high harmonics has been keenly pursued to produce high-frequency, coherent radiation sources. For the case of unmagnetized plasma,  theoretical analysis for a high harmonic generation was first given by \cite{margenau1948theory}. Experimentally, under a dc magnetic field, harmonic generation in microwave-induced gas-discharged plasma had been reported by \cite{hill1959harmonic}. Over the decades' many authors have contributed to this pursuit \cite{sodha1965third,sodha1970theory,basov1979second,gibbon1997high}.  Theoretical studies of a third harmonic due to the interaction of a microwave with plasma in a nonsteady state have been described by \cite{tripathi1971non}. High Harmonics in the reflected radiation of intense electromagnetic wave interaction with plasma slab by relativistic oscillating mirror model was first proposed by \cite{bulanov1994interaction} and the selection rules for identifying the polarization of reflected harmonics from an overdense plasma has been discussed by \cite{lichters1996short}. Analytical investigation of Second harmonic generation in plasma-filled cylindrical waveguide interaction with the High Power Microwave (HPM) has been done by \cite{fu2008harmonic}.   
In this work, we have explored the question of producing harmonics by the EM wave in a magnetized plasma through PIC simulations using EPOCH. The magnetic field has been chosen to be along the EM wave's propagation direction, termed the R-L mode geometry. Though we have chosen MW parameters for this study, it is equally applicable to possible laser parameters, which we have also listed. The question of harmonic generation in the context of X and O  mode geometry (for which the external magnetic field is applied along the laser magnetic and electric field, respectively) has been studied earlier \cite{maity2021harmonic}. In contrast to previous X and O mode configurations, we observe the formation of only odd harmonics. The dependence of the efficiency of harmonic generation on the applied magnetic field and other system parameters has also been investigated.

   The paper is organized as follows: Section \ref{sec:SimulationDetails} describes simulation geometry.  The  parameters used in the simulation are also  mentioned. In section \ref{sec:observations} we have provided a comprehensive analysis of various simulations performed for different ranges of parameters. We have illustrated various possible cases by varying the external magnetic field. The theory behind the generation of harmonics  is also described in this section. The effect of  the external magnetic field, EM wave polarization, and its intensity  on the generation of harmonics have been investigated. The analytical calculation  for the process of harmonic generation has been provided  in Appendix \ref{appA}. Finally, we have summarized our studies in  section \ref{sec:summary}. 
   
   \begin{figure}
  \centering
  \includegraphics[width=6.0in]{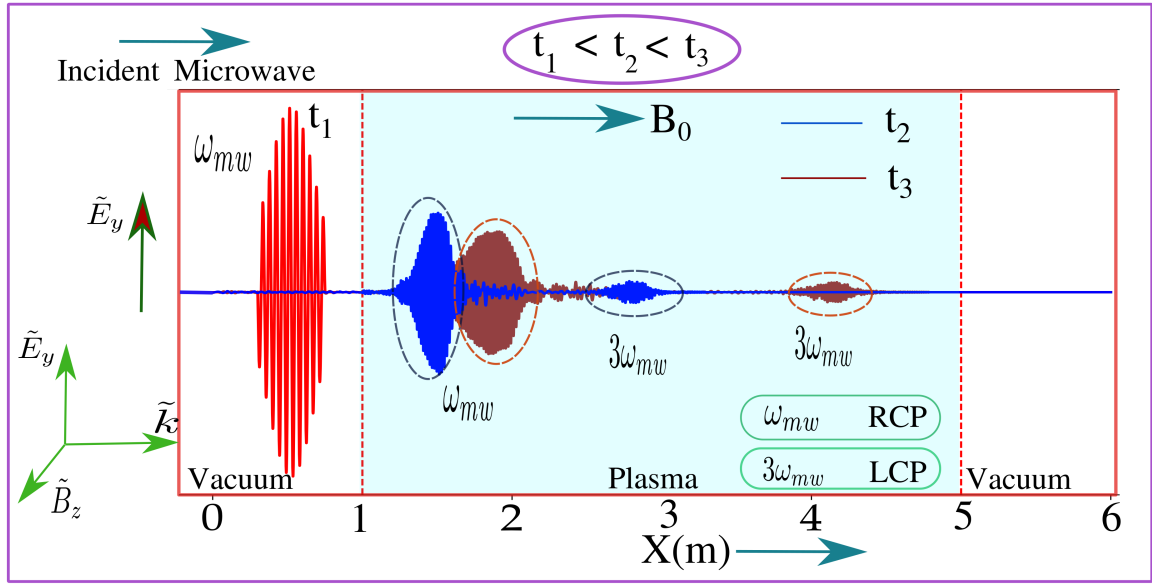}
  \caption{A schematic of our simulation box is shown in figure. We have carried out one dimensional PIC simulation. Here, external magnetic field $B_0$ is applied along direction of propagation of linearly polarized microwave $(\omega_{mw})$ (along $\hat x$ direction). As the HPM interacts with the plasma at surface we observe it excites higher harmonics of different polarisation (L-polarized) while passes the fundamental R-wave. All the higher harmonics satisfies the dispersion relation.}
\label{fig:schematic}
\end{figure}
\section{Simulation details}\label{sec:SimulationDetails}
The interaction of high-power microwave pulse with magnetized plasma has been studied with the help of  a one-dimensional particle-in-cell (PIC) simulation. The simulation has been carried out using  EPOCH 4.17.16 PIC code \cite{arber2015contemporary,bennett2017users}. The schematic of the simulation geometry is shown in figure \ref{fig:schematic}. We have divided our simulation box of 6 meters into 120,000 grids for which  $dx=50\mu m$. The plasma boundary starts from $x = 1 m$ and extends up to $x = 5$ meters. We have considered fully ionized homogeneous electron-proton plasma in our simulation box, where the ion-electron mass ratio is $1837.2$. The number density of electrons and ions is chosen to be constant in the plasma region $n_{e,i}=2\times10^{18} m^{-3}$.   In terms of skin depth thus the grid size corresponds to $dx=0.013c/\omega_p$. The number of macro particles per cell has been chosen to be $20$. We have considered a short pulse microwave of wavelength $\lambda=44.88 mm$ with a pulse width equal to $10$ wavelengths. Profile of laser pulse is chosen Gaussian in time with a peak intensity of  ($I=5.86\times 10^{11} W \ m^{-2}$). The transverse electric field of microwave $\tilde{E}_{mw}$ is along $\hat y$ direction and the oscillating magnetic field $\tilde{B}_{mw}$ of the wave is  in $\hat z$ direction. Microwave enters in simulation box from the left direction at $t=0s$.

\begin{table}
	\caption{Simulation parameters are shown here in normalized as well as in corresponding SI units}
	\label{table:simulationtable}
	\begin{center}
		\begin{tabular}{|c|c|c|c|}
			\hline		
			$$\textbf{ Parameters}$$       & $$ \textbf{ Normalized values}$$   &   $$  \textbf{ Microwave System}$$ & $$  \textbf{Laser System}$$ \\[3pt]
   \hline	
			\multicolumn{4}{|c|}{\textbf{Microwave/Laser Parameters}} \\
			\hline
                 Frequency($\omega_{mw}$)  & 0.53$\omega_{pe} $ & $4.2\times 10^{10}rad \ s^{-1}$ & $0.2\times 10^{15}rad \ s^{-1}$ \\ 
                 \hline
                 Wavelength($\lambda_{mw}$) & 12.1$c/\omega_{pe}$ &  44.88 $mm$  &$ 9.42 \mu m $\\
                 \hline
                 Intensity ($I_0$)           & $a_0=0.29$        &  $5.86\times 10^{11} W \ m^{-2}$ & $1.33\times 10^{19} W \ m^{-2}$\\
			\hline	
			\multicolumn{4}{|c|}{\textbf{Plasma Parameters}} \\
			\hline
			 Density$(n_{e,i})$   & 1  & $2\times 10^{18}m^{-3}$ &  $4.47\times 10^{25} m^{-3}$\\
              \hline
              ($\omega_{pe})$   & 1 & $7.973\times10^{10}rad \ s^{-1}$ & $3.77\times 10^{14} rad \ s^{-1}$ \\
              \hline
              ($c/\omega_{pe})$ & 1 & 3.7 mm & $0.79 \mu m$\\
		   \hline
		\end{tabular}	
	\end{center}
\end{table}
External magnetic field $B_0$ is chosen to be along propagation direction $\hat x$ for the R-L mode configuration. In the absence of an external magnetic field, microwaves cannot penetrate inside the  plasma as it is overdense $\omega_{mw}<\omega_{pe}$. We have chosen two different values of applied external magnetic fields. This ensures that in one case the incident laser is in the stop band of both L and R waves and the EM wave does not penetrate the plasma. Whereas for the second choice, while  the L mode is in the stop band  the R mode is in the pass band. 
Boundary conditions are taken as absorbing for fields as well as particles.    Since the ion-electron mass ratio is high which makes electron cyclotron frequency to be quite high ($\approx 1837.2$)  compared to the ion cyclotron frequency ($\omega_{ce}>>>\omega_{ci}$). At the  chosen MW frequency, only the lighter  electron species will  respond. 

\section{Observations and Discussion}\label{sec:observations}
  We report four cases of parameters for which simulations have been carried out in Table \ref{tab:Magneticfieldparameters}.  The polarization of the microwave/EM wave has also been shown in Table \ref{tab:Magneticfieldparameters}. Microwave/EM  frequency in all these cases has been chosen as $0.51\omega_{pe}$, and its intensity is chosen to be $a_0 = 0.29$ for all the studies.   We have observed that even if the fundamental frequency of microwave lies in a cutoff of both R $\&$ L-modes, some traveling structures (EM waves) are still present inside the bulk plasma. We have analyzed these EM disturbances using various diagnostics for each case.

\begin{table}
	\caption{List of simulations performed for linear and circularly polarized wave with applied external magnetic field.}
	\label{tab:Magneticfieldparameters}
	\begin{center}
		\begin{tabular}{|c|c|c|c|c|c|c|}
			\hline		
			 \textbf{Case}  & \textbf{Polarisation} & $\textbf{( MW )} B_0$ (T)  & $\textbf{ (Laser)} B_0$ (kT) & $\omega_{ce}$ ($\omega_{pe}$) & \textbf{R mode} & \textbf{L mode}  \\[3pt]
			\hline
                 A & linear & 0.1775  &0.839 &  0.39 & cut off & cut off  \\
              \hline
                 B & linear & 0.355  &1.68 & 0.78 & \color{red}pass band & cut off \\
              \hline
                 C & RCP    & 0.355  &1.68 &  0.78 &\color{red} pass band & cut off \\
             \hline
                 D & LCP    & 0.355  & 1.68 &  0.78 & \color{red}pass band & cut off \\
		   \hline
		\end{tabular}	
	\end{center}
\end{table}

\begin{figure}
  \centering
  \includegraphics[width=6.0in]{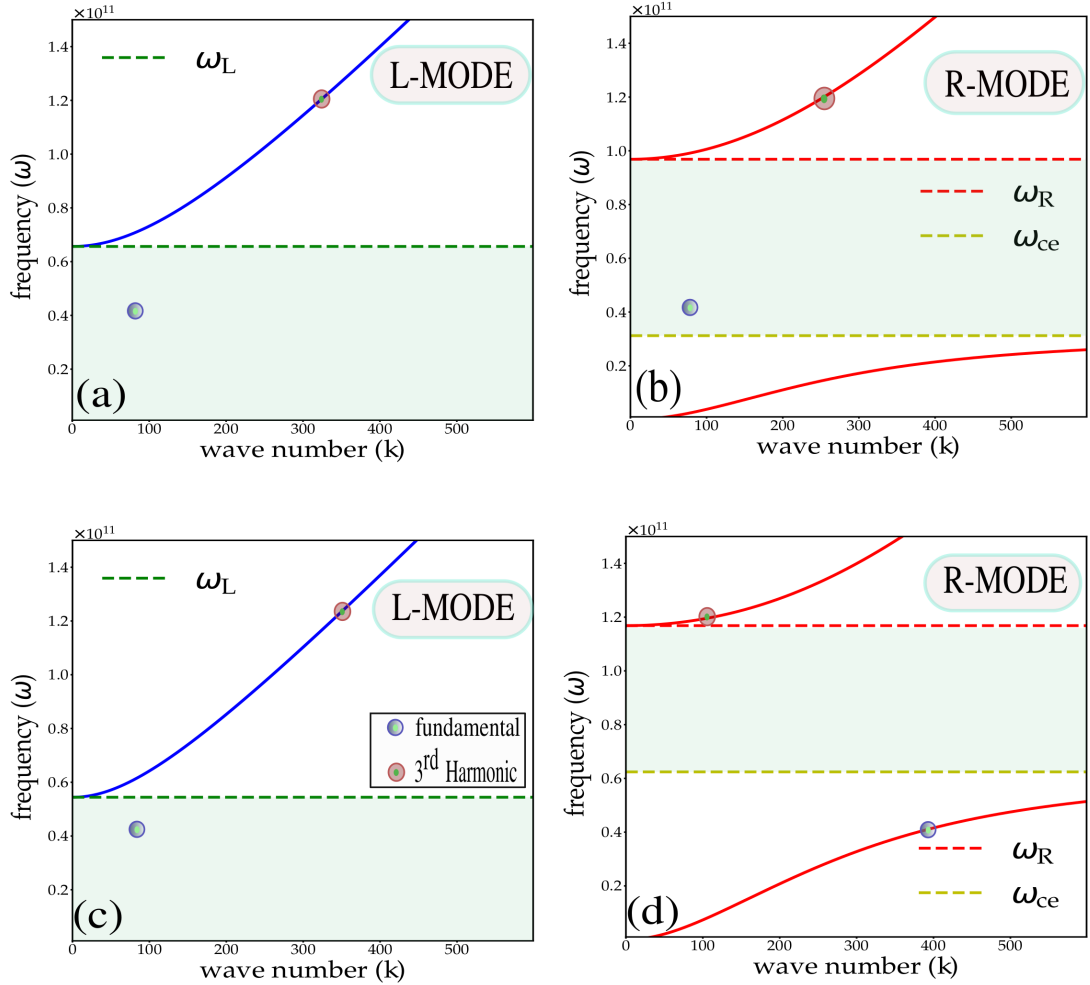}
  \caption{Dispersion curves of L and R mode corresponding to case A ($B_0=0.1775T$) shown in figure (a) and (b) respectively.  Similarly, figure (c) and (d) illustrates dispersion curves of L and R mode for case B ( $B_0=0.355T$)  \cite{chen1984introduction}.}
\label{fig:Dispersion curve}
\end{figure}

\begin{figure}
  \centering
  \includegraphics[width=6.0in]{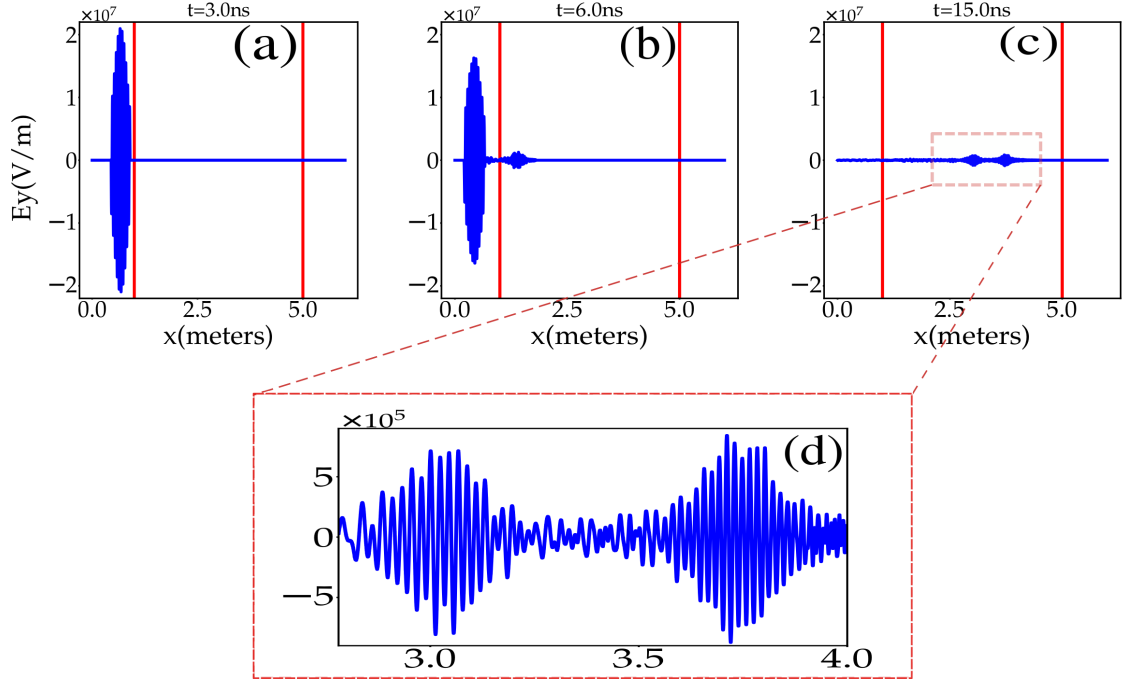}
  \caption{Time evolution of interaction of incident EM wave with magnetized plasma is shown in above figures $(a-c)$ at $t=3, 6, 15 ns$. fig-$(d)$ shows zoomed version of $E_y$ component of transmitted structures.   }
\label{fig:time_evol_A}
\end{figure}

\begin{figure}
  \centering
  \includegraphics[width=6.0in]{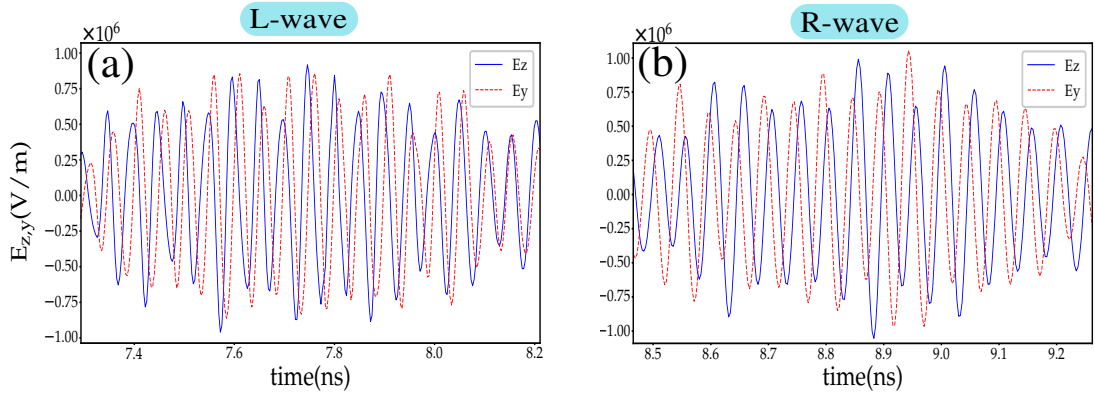}
  \caption{Time evolution of the electric field component along y and z direction is plotted here for fast $(a)$ and slow $(b)$ wave inside plasma }
  \label{fig:polarisation_t}
\end{figure} 

\begin{figure}
    \centering
    \includegraphics[width=6.0in]{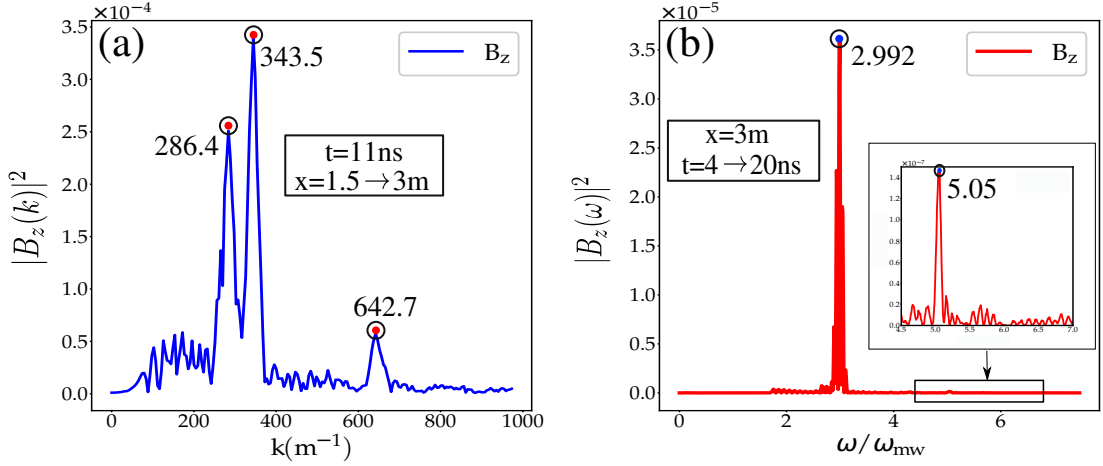}
    \caption{Fast Fourier transform was performed in time $(a)$ and space $(b)$ to measure the corresponding frequency and propagation vector for the two travelling structure in magnetized plasma. }
    \label{fig:FFTcaseA}
\end{figure}

\begin{figure}
\centering
  \includegraphics[width=6.0in]{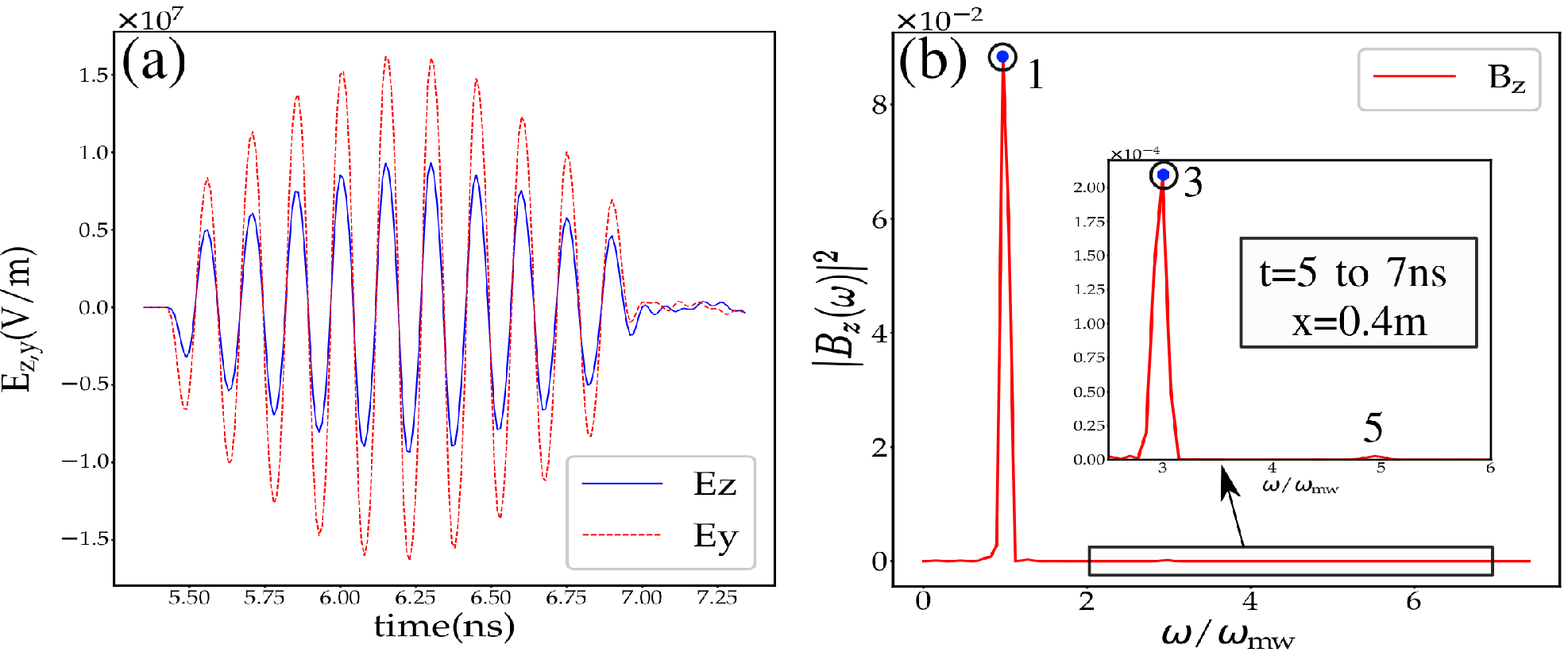}
  \caption{figure $(a)$ shows the time evolution of electric field along y and z direction of reflected wave in vacuum region. While, figure $(b)$ depicts time FFT of $|B_z(\omega)|^2$ in vacuum region. }
\label{fig:polarisation_r}
\end{figure} 

 \subsection{Case A}
We choose the value of the applied external magnetic field of $0.3 m_ec\omega_{pe}e^{-1}$, which corresponds to $B_0=0.1775T$ for the microwave-plasma interaction studies. The direction of the magnetic field is chosen to be along the propagation vector of the microwave in the plasma. The microwave frequency has been chosen greater than electron cyclotron frequency, i.e. $\omega_{mw}>\omega_{ce}>>>\omega_{ci}$. Here $\omega_{cs}$,  represents ion and electron cyclotron frequency for $s=i,e$, respectively. 
Also, for this magnetic field $\omega_{mw}<\omega_{R,L} $, where $\omega_{R,L}$ represents right and left-hand cut-off frequency. So for this case, the dispersion relation governed by the magnetized plasma medium is given by:   \cite{chen1984introduction}. 
\begin{equation}\label{eq:1}
    \frac{c^2k^2}{\omega^2}=1-\frac{\omega_{pe}^2/\omega^2}{1\mp\omega_{ce}/\omega}
\end{equation}
Where '$\mp$' signs correspond to R-mode/L-mode, respectively. Now for this case, the equation (\ref{eq:1}) tells us that incident EM wave frequency falls under the cut-off region of both R and L modes. This can be observed from the dispersion relation shown in figure \ref{fig:Dispersion curve}(a,b). Thus, for this particular case, one expects the incident EM wave to reflect from the plasma boundary. There would be no propagation inside the plasma. The simulations, however, show something different. The snapshots at various times of incoming EM wave interacting with magnetized plasma for this case have been shown in figure \ref{fig:time_evol_A}. The microwave/EM pulse enters from the left side of the simulation box. Figure \ref{fig:time_evol_A}$(a)$ at $t= 3 ns$ shows it is still in the vacuum region. Later, it interacts with the plasma boundary at $x=1 m$. At $t= 6 ns$ we can see the two wave pulse in the simulation box (figure \ref{fig:time_evol_A}$(b)$) transmitted and reflected pulse. While the reflected amplitude of the wave is high, one also observes a small amplitude of a wave that seems to propagate inside the plasma. Subsequently, at $t=15ns$, the EM wave in bulk is also observed to separate in two distinct pulses. A zoomed view of these two structures has also been shown in figure \ref{fig:time_evol_A}$(d)$. 
    
\indent To understand the behavior of the transmitted pulses, we carry out various diagnostics. It is observed that both $y$ and $z$ components of the electric field are finite for these pulses, unlike the incident linearly polarised EM pulse.   The time evolution of the electric field vectors $E_y, E_z$ at a fixed location of $x=3m$ for the two pulses has been shown in figure \ref{fig:polarisation_t}. From the phase difference between $E_y $ and $E_z$, one infers that the faster-moving pulse is left circularly polarized (L-wave). On the other hand, the slow-moving wave is found to be right circularly polarized (R-wave). The  Fast Fourier transform in time at the same location $x=3 m$ (from $ t= 4$ to $20 ns$) for which these pulses cross this particular point has been obtained and shown in figure \ref{fig:FFTcaseA}$(b)$. The dominant peak is near the third harmonic, and the other smaller peaks occur at higher odd multiples of the incident wave frequency ($\omega_{mw}$). We also performed the space FFT analysis in figure \ref{fig:FFTcaseA}$(a)$ at $t=11 ns$  when the two pulses extend from $ x=1.5$ to $3m$ inside the box. In this case the dominant peaks occur  at $k= 286.4 m^{-1}$ and $343.5 m^{-1}$. The observed values of dominant peaks in $\omega$ and $k$ satisfy dispersion curves of R and L mode shown in figure \ref{fig:Dispersion curve}$(a,b)$ for the applied external magnetic field of $(0.39m_ec\omega_{pe}e^{-1})$. We similarly analyze the properties of the reflected wave. Figure \ref{fig:polarisation_r}$(a)$ shows that the reflected wave also has both $E_y$ and $E_z$ components, but both are in phase. Thus the reflected wave remains linearly polarized, but its plane of polarization gets rotated along the yz-plane.

These observations thus show that though the fundamental frequency lies in the stop band (for both L and R waves) and, therefore, cannot propagate inside the plasma, high harmonics get generated at the vacuum plasma surface and propagate inside the plasma.  The $3^{rd}$ harmonic is the most dominant frequency generated.  Time FFT spectrum in figure \ref{fig:polarisation_r}$(b)$ depicts that there are also higher odd harmonics present in the transmitted region.  

\subsection{Case B}
   We next choose a case for which the strength of the external magnetic field is doubled to $0.355T$ ($0.78 m_ec\omega_{pe}e^{-1}$). For this case (B) microwave frequency holds  $\omega_{ce}>\omega_{mw}>\omega_{ci}$. Subsequently, in this case, microwave frequency falls under the range $\omega_{ce}/2<\omega_{mw}<\omega_{ce}$, where the group velocity decreases while the phase velocity of the incident wave increases. So, the microwave frequency lies in the pass band of the R mode dispersion curve figure \ref{fig:Dispersion curve}$(c)$ 
   However, in the cut-off region of the L-mode dispersion curve shown in figure \ref{fig:Dispersion curve}$(c)$ (shown by blue dots). In this case, a part of the incident EM wave travels inside the plasma as a Right circularly polarized wave at the fundamental frequency. Higher odd harmonics also get generated, which propagate inside the plasma. It should be observed from the dispersion curve (for the applied magnetic field of  this particular case) shown in figures \ref{fig:Dispersion curve}$(c,d)$ depicts that $3^{rd}$ harmonic of both R and L polarization lie in the pass-band. One would, therefore, expect the $3^{rd}$ harmonic of both polarization to get generated like the previous case and propagate inside the plasma. 
   
   \begin{figure}
  \centering
  \includegraphics[width=6.0in]{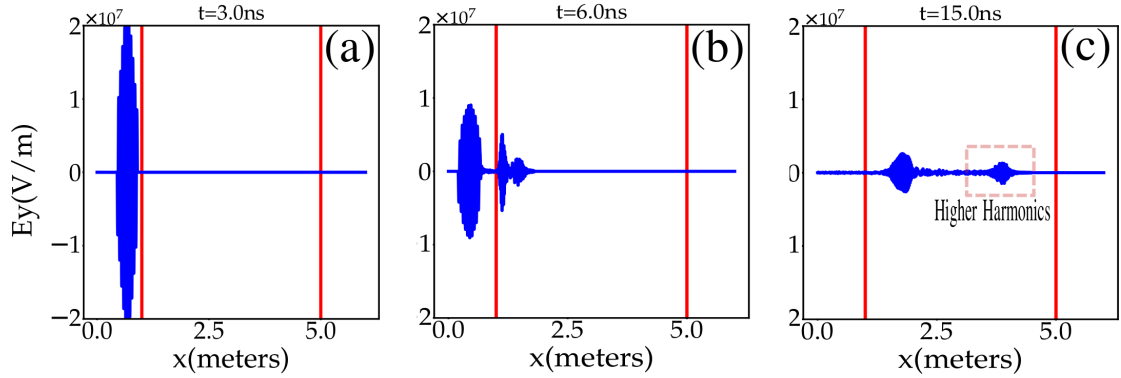}
  \caption{ Time evolution of interaction of incident EM wave with magnetized plasma is shown in figures (a-c) at t=3, 6, 15 ns for the Case B. There are two travelling structures inside the magnetized plasma has been detected.}
\label{fig:time_evol_B}
\end{figure}

\begin{figure}
  \centering
  \includegraphics[width=6.0in]{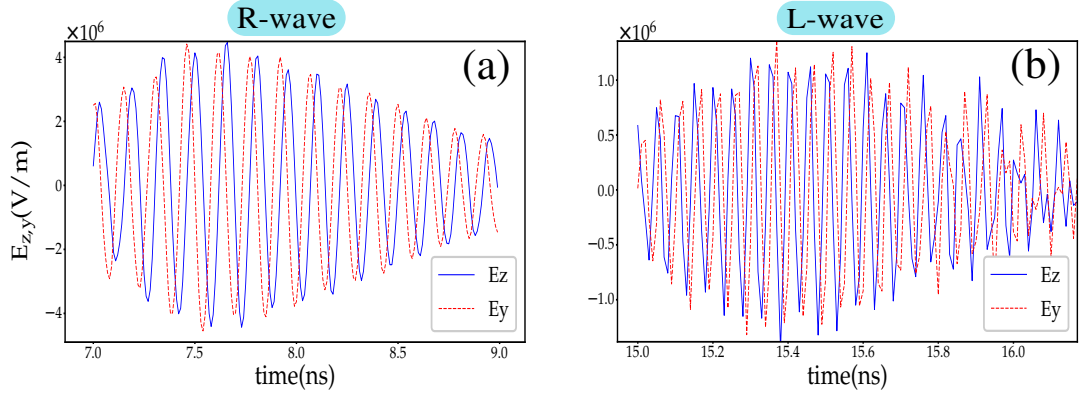}
  \caption{ Subplots of Electric field components along y and z direction has been plotted with time. Figure (a) correspond to the slow travelling wavepulse, while figure (b) corresponds to the fast travelling wavepulse.  }
\label{fig:polarisation_harmonic}
\end{figure}
   \begin{figure}
    \centering
    \includegraphics[width=6.0in]{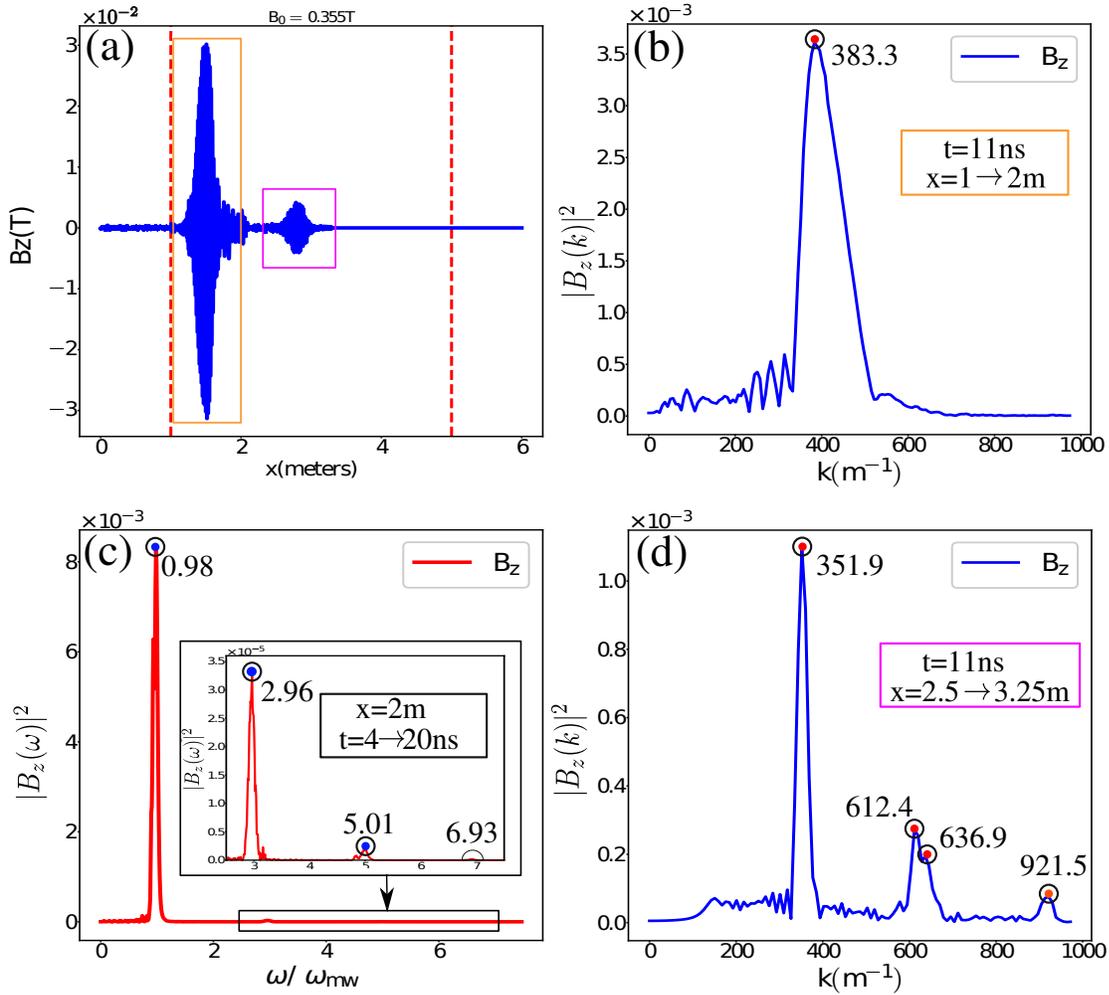}
    \caption{ Figure (a) shows the snapshot of transverse magnetic field ($B_z^{(mw)}$ of microwave inside magnetized plasma at t=11 ns. Figure (b) and (d) depicts space FFT associated to each of these pulses, and figure (c) demonstrates time FFT at $x=2$ meters in space from time period of $t= 4$ to $20 ns$.}
    \label{fig:FFTcaseB}
\end{figure}

    We have shown the time evolution of incident microwaves in this case. Figure \ref{fig:time_evol_B}$(a)$ depicts that at $t= 3ns$ Gaussian wave pulse of the incoming microwave reached the vacuum plasma interface. After $t=6ns$, the microwave entered the plasma, and a reflected wave from the plasma boundary could be seen in figure  \ref{fig:time_evol_B}$(b)$. Later, after $t=15 ns$ reflected wave has left the simulation box, two clear, distinct wave pulses are observed in magnetized plasma. The slow traveling wave-pulse in figure \ref{fig:time_evol_B}$(c)$ turns out to be the fundamental R-mode as indicated by the dispersion curve in figure \ref{fig:time_evol_B}$(d)$. Besides that, we observe only one fast-moving wave-pulse instead of two, as one would expect from the dispersion curve in figures \ref{fig:Dispersion curve}$(c,d)$. The polarization of these waves has been depicted in figure \ref{fig:polarisation_harmonic}, which describes the polarisation of the slow-moving waves as the R-wave and the fast-moving wave as an L-wave. Figure \ref{fig:FFTcaseB} (a) shows the simulation snapshot at $t=11 ns$. The time FFT for the time window of $t=4$ to $20 ns$ has been evaluated until the whole wave pulse passes through from x=2 m. It shows the fundamental peak at $\omega=0.98\omega_{mw}$ and small peaks at  $\omega/\omega_{mw}= 2.96, 5.01, 6.93 $. Figures \ref{fig:FFTcaseB}$(b,c)$ illustrates the space FFTs for the two wave packets shown in (a). There corresponding k values for slow wave-packet at $x= 1$ to $2$ metres is found at $k= 383.3 m^{-1}$. While for the fast moving wave-pulse the maximum at $k=351.9 m^{-1}$ and further small peaks at $k= 612.4, 636.9 m^{-1}$ and $921.5 m^{-1}$ has been observed. This ensures that odd harmonics have excited in plasma along with the fundamental R-wave. Interestingly, the $3rd$ harmonic of only the L-wave got excited in the plasma. 
  
\begin{figure}
  \centering
  \includegraphics[width=5.0in]{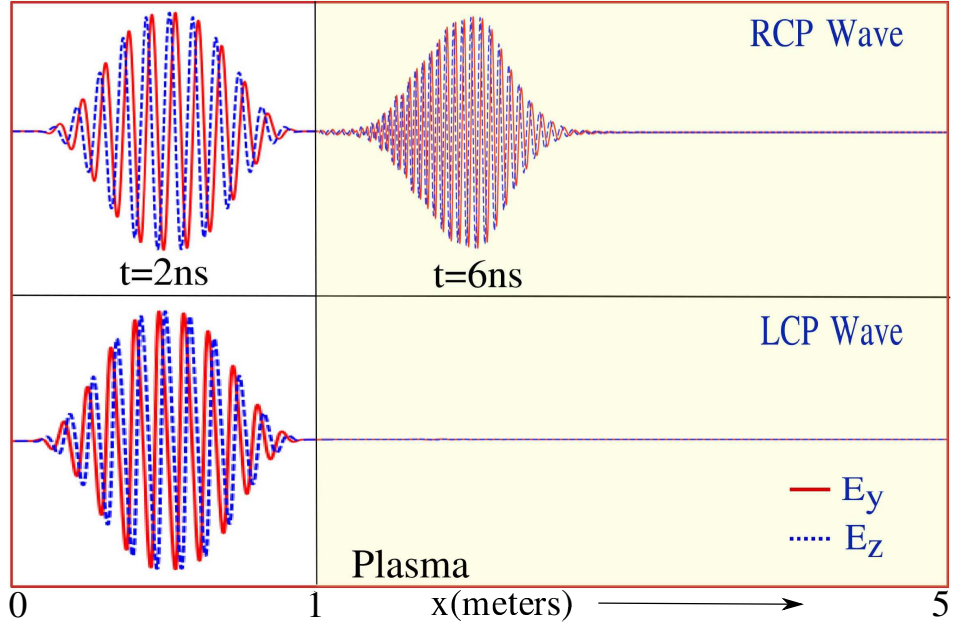}
  \caption{Figure depicts time evolution of RCP and LCP wave encountering RL-mode in magnetized plasma at external magnetic field $B_0=0.355T$. At $t=2 ns$ profile of RCP and LCP wave has shown. Later in time, at $t=6 ns$ following the dispersion relation RCP wave pass through plasma and LCP wave reflects from surface.}
\label{fig:polarisation}
\end{figure}

\subsection{Case C and D}
 A circularly polarized microwave (RCP and LCP) has been made to fall on the magnetized plasma. Figure \ref{fig:polarisation} shows when incident right circularly polarized microwave of frequency ($\omega=4.2\times 10^{10}rad/s$) is set to fall on plasma medium only fundamental R-wave travels with the finite group velocity. Similarly, when the left circular polarized microwave passes through the magnetized plasma, it follows the L-mode dispersion relation, and all the incident microwave gets reflected from the plasma boundary. These two cases reveal that no higher harmonics get excited for a circularly polarized light in the magnetized plasma. This indicates that the polarisation of the wave also has an essential role in the generation of high harmonics. The reason behind this will be covered in the following section.

\section{Symmetric Generation of Harmonics at plasma boundary }
Excitation of only odd harmonics in RL-mode ($ \tilde k||B_0$) configuration could be explained by invoking the symmetry of our chosen simulation geometry. 
The incoming EM wave with arbitrary polarisation propagating along $x$ direction has associated Electric ($\tilde E_{mw}$) and Magnetic ($\tilde B_{mw}$) fields that are in the  $\hat y$ and $\hat z$ plane orthogonal to each other. A detailed perturbative expansion has been carried out in the Appendix.  
This geometry depicts that in the linear regime electrons at the surface of the vacuum-plasma interface,  in response to the external oscillating Electric field ($\tilde E_{mw}$) start oscillating in yz-plane with their quiver velocities $v_y^{(1)}$ and $v_z^{(1)}$ with EM wave frequency ($\omega_{mw}$) according to equation (\ref{eq:velocity1st}). Consequently, this will excite the surface current $\tilde J_{ey}(\omega_{mw})$ and $\tilde J_{ez}(\omega_{mw})$. Electrons oscillating in the yz-plane further couples with the oscillatory magnetic field ($\tilde B_{mw}$) of the microwave, which is in yz plane through Lorentz force ($~\tilde v^{(1)} \times \tilde B_{mw} exp(-2i\omega_{mw}t)$). Consequently, This will excite second harmonic ($\tilde v_x^{(2)}$) along the $\hat x$ direction. Thus, the surface current will be excited only along the x-direction in the second harmonic. Which could be described by the equation (\ref{eq:currrentdensity}).
\begin{equation}
    \tilde{J_{ex}}^{(2)}=-\frac{n_ee^3}{2m^2c}\frac{E_{mw0}^2}{(\omega_{mw}^2-\omega_c^2)}(1+\alpha^2)exp(-2i\omega_{mw}t)
    \label{eq:currrentdensity}
\end{equation}
\indent It is interesting to note that for circular polarization since  $\hat y$ and $\hat z$-component of $\tilde B_{mw}$ will be equal and have phase difference with each other ($\alpha =\pm i $) that makes ( $J_{ey,ez}(2\omega)\times B_{l(y,z)}$ terms in second harmonic to vanish,  illustrating from the equation-(\ref{eq:currrentdensity}) that for circular polarisation current density in second harmonic cannot be excited. This is why further coupling to generate $3^{rd}$ or higher harmonic is impossible for circularly polarized waves. This was confirmed earlier by simulations in cases C and D. The geometry being 1-D with variations along $x$, this second harmonic current along $x$ does not generate any radiation.
 \begin{figure}
  \centering
  \includegraphics[width=6.0in]{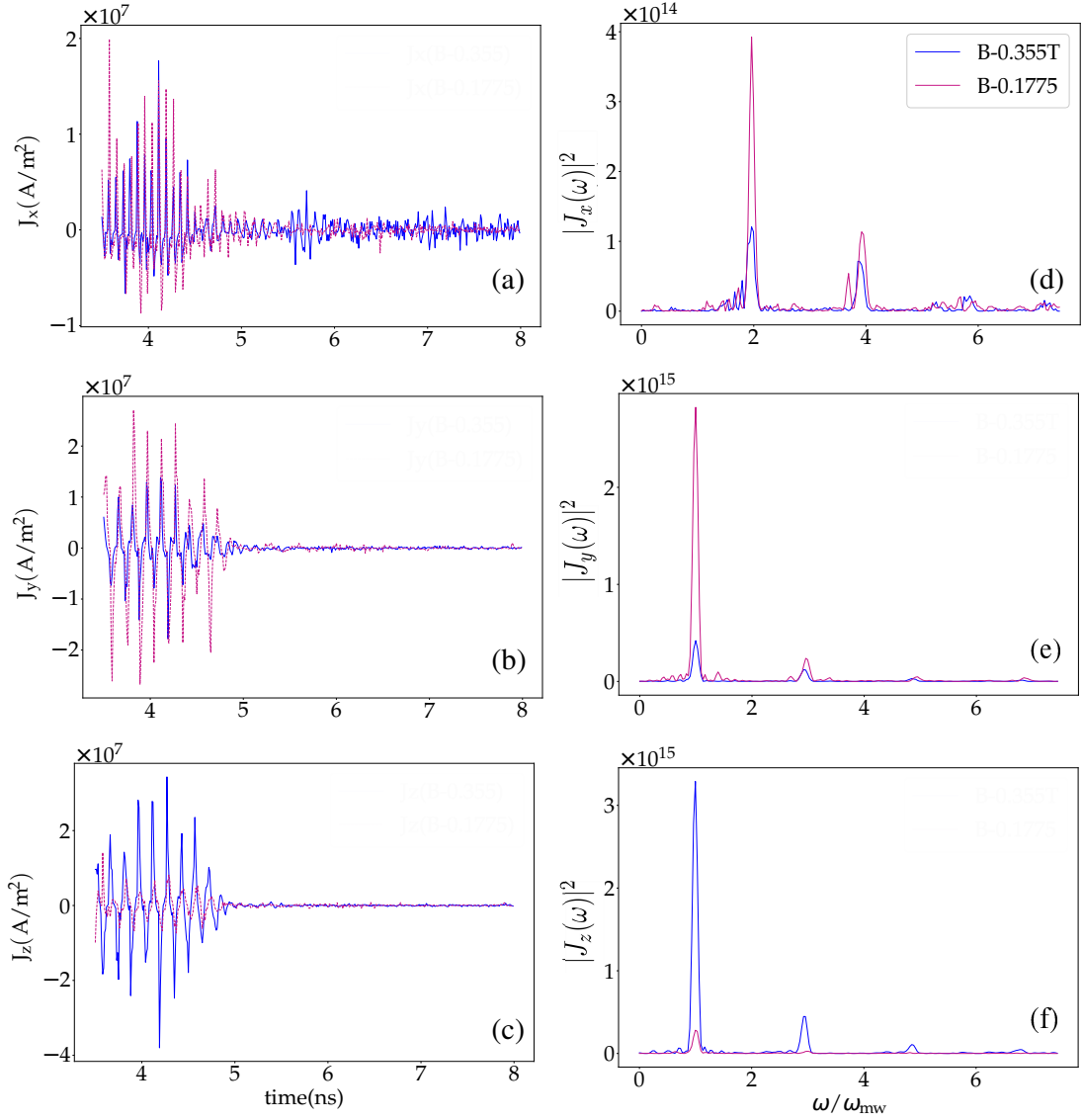}
  \caption{Figure $(a,b,c)$ elucidate time evolution of current density along $\hat x$,  $\hat y$ and  $\hat z$ direction at surface of vacuum plasma interface for the cases A and B. Figure $(d,e,f)$ describes the time FFT's corresponding to these current densities at the surface.}
\label{fig:surfacecurrent}
\end{figure}

\indent  Furthermore, since current density in second harmonic will be finite for linear polarized light ($\alpha=0$ in \ref{Elec_field},\ref{Mag_field}) only the  electrons wiggles with frequency ($2\omega_{mw}$) along x-direction at the surface of plasma and couples with  oscillatory magnetic field ($\tilde B_{mw}$)  through the Lorentz force ($~\tilde v^{(2)}\times \tilde B_{mw} exp(-i3\omega_{mw}t)$) again. By symmetry, it will excite the third harmonic current in yz-plane. Thus, surface current in third harmonic will be along $\hat y$ and $\hat z$ direction. Thus one will obtain  odd harmonic radiation (Electromagnetic modes) from this surface current.  In the other geometry of  X  and O-mode  which was reported earlier by \cite{maity2021harmonic} even harmonics were also got generated.

\indent In our simulation we have been able to observe up till $7^{th}$ harmonic in FFT spectrum. Figure (\ref{fig:surfacecurrent}) depicts surface current generated at the vacuum plasma boundary along each directions for the cases A and B. Figure (\ref{fig:surfacecurrent})$(a)$ shows variation of surface current for both cases along x-direction with time (from $t=3.5$ to $8 ns$) up till the incoming microwave pulse interacts with the vacuum plasma interface. Time Fast Fourier transform (FFT) of the interaction has been shown in figure \ref{fig:surfacecurrent}$(b)$. The FFTs clearly reveals that current density corresponding to even harmonics ($\omega=2\omega_{mw},4\omega_{mw},6\omega_{mw}$) have been excited at the surface. Since the even harmonics cannot propagate inside the plasma, they only exist at the surface and further excite odd harmonics which can propagate through plasma according to the specific dispersion relation. Similarly, figure \ref{fig:surfacecurrent}$(b,c$) shows surface current density excited along $\hat y$ and $\hat z$ direction. Figure \ref{fig:surfacecurrent}$(e,f)$ illustrates time FFT in y and z direction respectively, which clearly shows that current densities of odd harmonic frequencies got excited at the surface. Here, we observed that for lower magnetic field (case A) current density is greater along $\hat y$ than $\hat z$ direction but as we doubled the magnetic field (case B), current density  become greater along $\hat z$ then $\hat y$ direction. Which reveals that for increasing the magnetic field plane of linearly polarized light is rotating in yz-plane.

\section{Physics of Propagation of Harmonics in RL-Mode Configuration ($\tilde k||B_0$)}
\begin{figure}
    \centering
    \includegraphics[width=6.0in]{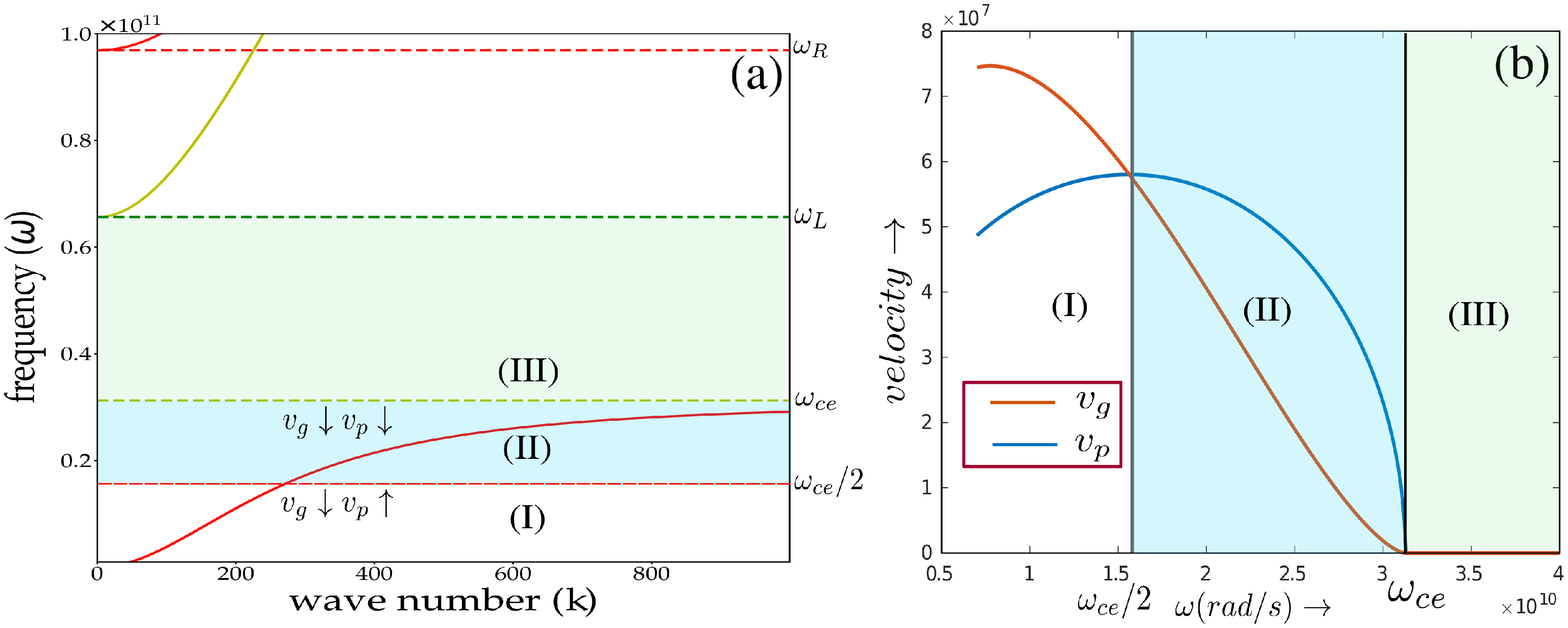}
    \caption{Figure demonstrates different regions of pass and cut off bands of R-mode depending on group and phase velocity of incoming EM wave in plasma under RL-mode configuration. For EM wave frequency($\omega_{mw}<\omega_{ce}/2 $) bounds region-I, subsequently ($\omega_{ce}/2 <\omega_{mw}<\omega_{ce} $) bounds region-II and ($\omega_{ce}<\omega_{mw}<\omega_{L}$) bounds region-III. }   \label{fig:dispersion_Rmode}
\end{figure}

To unravel the physics of propagation of harmonics in magnetized plasma along RL-mode geometry, we have considered a fully ionized cold electron-proton plasma in the present scenario. Since the microwave frequency ($\omega_{mw}$) is much higher compared to ion response time scales ($\omega_{pi}$,$\omega_{ci}$) in magnetized plasma. So they do not respond to the interaction at the surface but merely provide a neutralizing background. 
Figure \ref{fig:dispersion_Rmode}$(a)$ demonstrates three regions in the RL-mode dispersion curve which depicts different characteristics of generated high harmonics when the incident electromagnetic wave has chosen to be in any of these three regions. We will discuss these three regions briefly to understand the physics of harmonics in RL-mode Geometry. 

\subsection{Region-I}
 \indent When the incoming microwave frequency is less than $\omega_{ce}/2$ then it falls under region-I. Equation-(\ref{eq:1}) could be solved to derive the group velocity ($v_g$) and phase velocity ($v_p$) of incoming EM wave in the R-mode. Which then given as, 
 \begin{equation}
 v_p= \frac{\omega}{k} =c\sqrt{\frac{\omega(\omega_{ce}-\omega)}{\omega_{pe}^2 +\omega_{ce}\omega -\omega^2}}
 \label{eq:v_p}
 \end{equation}

\begin{equation}
 v_g=  \frac{2c^2 k(\omega_{ce}-\omega)}{c^2 k^2+\omega_{pe}^2 -3\omega^2 +2\omega_{ce}\omega}
 \label{eq:v_g}
\end{equation}
 Figure \ref{fig:dispersion_Rmode}$(b)$ depicts the variation of the group ($v_g$) and phase ($v_p$) velocity with incident EM wave frequency. Figure \ref{fig:dispersion_Rmode}$(b)$ describes that in the region for $\omega < \omega_{ce}/2$, with increasing the incoming wave frequency, group velocity decreases whereas the phase velocity of EM wave increases. Also, when the applied EM wave frequency is less than $\omega_{ce}/2$, the time scale of the incoming em wave ($\omega_{mw}^{-1}$) would be more significant compared with the cyclotron gyration frequency of electrons ($\omega_{ce}^{-1}$). Because of these circumstances, the surface interaction of EM waves with magnetized plasma cannot excite higher harmonics in the magnetized plasma. Furthermore, equation (\ref{eq:vel_harmonic}) in Appendix \ref{appA} illustrate that the denominator of quantity $v_{x}^{(2)}$ would become higher, which in turn makes surface currents ($J_{ex}^{(2)}$) in second harmonic to be lesser and coupling further to excite higher $3^{rd}$ harmonic would become rather low in efficiency as indicated by the denominator in equation (\ref{currentdens}). Therefore, in this region harmonic generation efficiency would be low. For this reason, harmonic generation was not observed in an earlier work by \cite{goswami2021ponderomotive}.
\subsection{Region-II}
 \indent Let us now understand the dynamics when the incident microwave frequency is chosen in the region $\omega_{ce}/2 < \omega < \omega_{ce}$. Equations (\ref{eq:v_p},\ref{eq:v_g}) illustrate that in this region group, and phase velocity of the transmitted wave in plasma decreases for increasing the incident EM wave frequency. The time scales of microwave and electron gyration frequency will also become closer as we move towards resonance frequency ($\omega_{ce}$). Equation (\ref{eq:vel_harmonic},\ref{currentdens}) will become significant to excite current densities in higher harmonics. That is why surface interaction will significantly excite higher harmonics in magnetized plasma.
 \indent Since the fundamental frequency of EM wave fall under the pass band of R-mode but in the cutoff region of L-mode. Consequently, when linearly polarized light is subjected to an incident on the vacuum plasma interface, it could be assumed as a superposition of RCP and LCP waves. In RL-mode configuration, RCP and LCP follow different dispersion relations and travel with different group velocities. In this case, the RCP wave will pass through the plasma while the LCP wave finds a reflecting boundary. Because of that RCP wave will not have enough time to excite the $3^{rd}$ harmonic at the surface. The LCP wave, which got stopped at the interface, will have sufficient time to excite the $3^{rd}$ harmonic of the L-wave. Therefore, the efficiency of $3^{rd}$ Harmonic of the L-wave is significantly higher than the R-wave in this region. This makes us unable to detect the $3^{rd}$ harmonic of the R-wave in the case of B, even when it lies within the pass band of the R-wave.
 
 \subsection{Region III}
 \indent In this region frequency of the EM wave falls within the cutoff band of both R and L-mode configurations. Figure (\ref{fig:dispersion_Rmode}) describes that boundaries of region-III are extended between $(\omega_{ce} ,\omega_{mw})$. When the incident linearly polarized EM wave has frequency falls within this region, it will be reflected from the vacuum plasma boundary. Linearly polarized waves are considered the superposition of RCP and LCP waves in a vacuum. The time scales of the interaction of both component of the wave is now equivalent, which is why if the $3^{rd}$ Harmonic of both R and L mode lies in the pass band ($3\omega_{mw}>\omega_{L, R}$) then we can observe odd harmonics of incident wave in this case. It has also been confirmed by our simulation performed in case A.   
\begin{figure}
  \centering
  \includegraphics[width=6.0in]{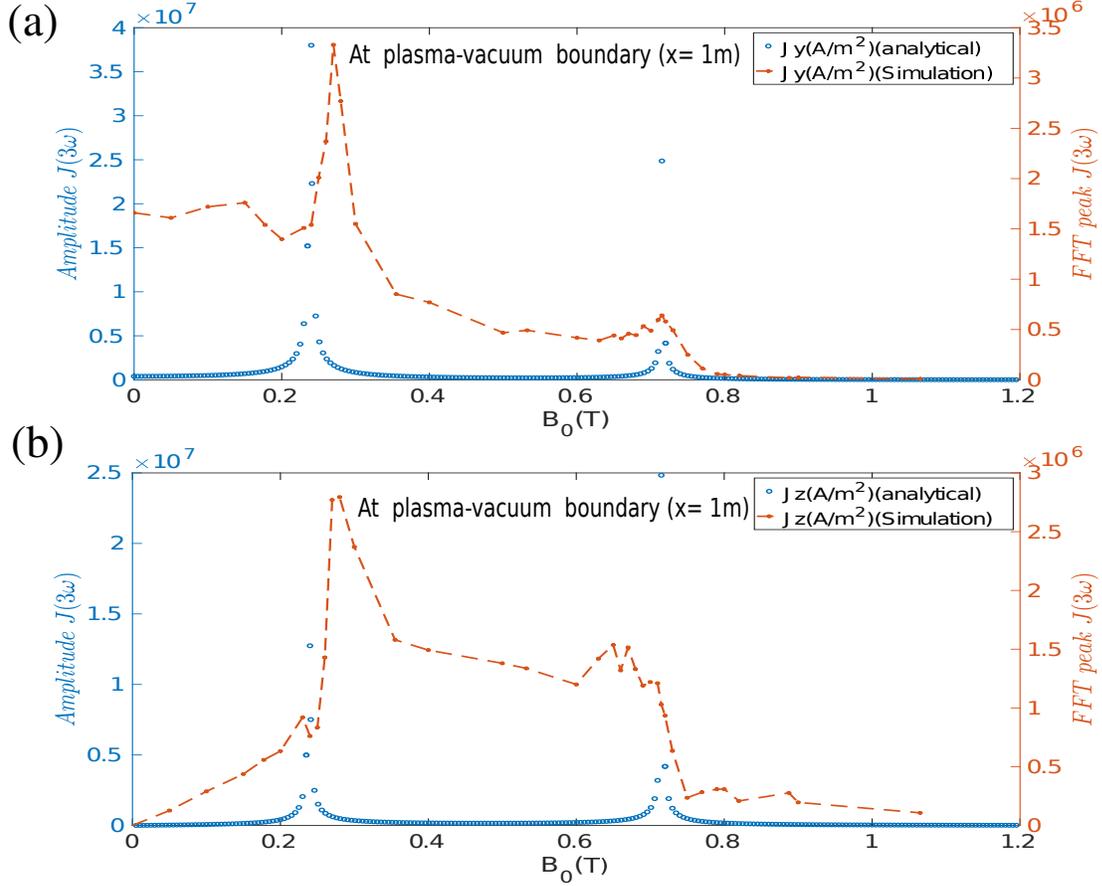}
  \caption{ Figure (a) and (b) depicts the variation of current density of third harmonic along y and z-direction at the vacuum plasma boundary for varying external magnetic field from $B_0$=$0$ to $1.2T$. The blue curve plots amplitude of the analytical expression-(\ref{currentdens}) and orange curve plot peak of $3^{rd}$ harmonic of FFT spectrum.}
\label{fig:JyJz}
\end{figure}
\section{Effect of magnetic field variation on harmonic generation}
We have observed that the excitation of harmonics and their efficiency depend on the region in which the frequency of incoming EM waves falls into the dispersion curves of RL mode. The externally applied magnetic field plays a crucial role in generating these harmonics. We have varied the strength of the magnetic field ranging from $B_0=0$ to $1.2 T$. This range of magnetic field will cover all three regions of the dispersion curve in figure (\ref{fig:dispersion_Rmode}) for incoming microwave frequency of $\omega_{mw}$. Thus to verify its effect on harmonics, we have plotted peak current density ($J_{ey}(3\omega_{mw})$ and $J_{ez}(3\omega_{mw})$ ) of $3^{rd}$ harmonic frequency from time FFT Spectrum along $\hat y$ and $\hat{z}$ directions as a function of the strength of applied external magnetic field. The analytical values have been obtained from the Appendix equation (\ref{currentdens}) using similar parameters as that of the chosen in simulation. It can be observed that the analytical current densities (\ref{currentdens}) follow a very similar trend with the external magnetic field, as the simulation shows. The location of the two peaks obtained from the perturbative approximate analytical expressions for ($\omega_{mw}=\omega_{ce}$) and ($\omega_{mw}=3\omega_{ce}$) are observed in simulation also. 
\begin{figure}
  \centering
  \includegraphics[width=6.0in]{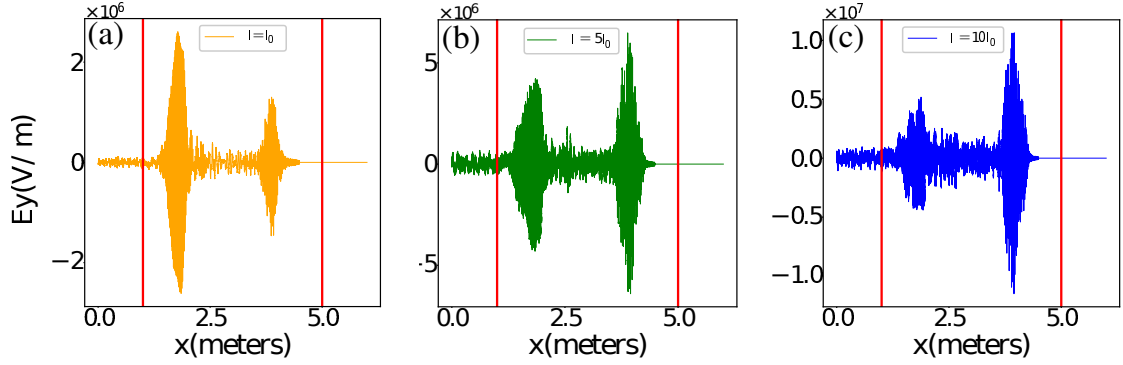}
  \caption{Figure (a,b,c) demonstrates the snapshot of  transverse Electric field ($E_y^{mw}$) of microwave in transmitted region for incoming microwave of intensities ($I_0$,$5I_0$, $10I_0$) corresponding to $I_0=5.857\times 10^{11}W/m^2$ at t= 15s, while the applied external magnetic field is $B_0=0.355T$.   }
\label{fig:Intensity_variation}
\end{figure}
\begin{figure}
  \centering
  \includegraphics[width=6.0in]{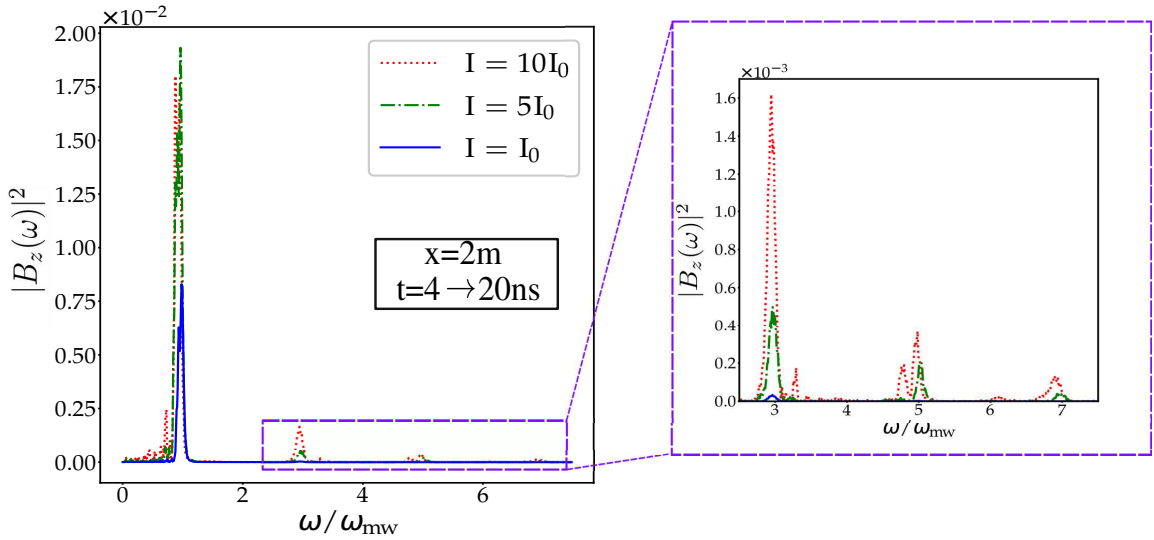}
  \caption{Figure demonstrates the time Fast Fourier transforms of transverse magnetic field ($B_z^{mw}$) of microwave in transmitted region for incoming microwave of intensities ($I_0$,$5I_0$, $10I_0$) corresponding to $I_0=5.857\times 10^{11}W/m^2$, while the applied external magnetic field is $B_0=0.355T$.   }
\label{fig:Intensity_fft}
\end{figure}
\begin{table}
        \centering
        \begin{tabular}{|c|c|c|c|c|}
            \hline
            Intensity & Harmonics ($n\omega_{mw}$) & {$\eta_{ref}$($\%$)} &            \multicolumn{2}{|c|}{$\eta_{trans}$($\%$)}    \\
            &  &  & \L-wave &  R-wave \\
            \hline
          $I=I_0$ & $3\omega_{mw}$ & $0.074$ & $0.343$ & $0$ \\    
          $(a_0=0.29)$ & $5\omega_{mw}$ & $6.997\times 10^{-4}$ & $0.0024$ & $0.027$ \\
            & $7\omega_{mw}$ & $0$ &\multicolumn{2}{|c|}{$0.0021$ } \\ 
        \hline
      $I=5I_0$ & $3\omega_{mw}$ & $0.355$ & $1.27$ & $0$ \\ 
         $(a_0=0.65)$ & $5\omega_{mw}$ & $0.0188$ & $0.0607$ & $0.33$  \\
         & $7\omega_{mw}$ & $0$ & \multicolumn{2}{|c|}{$0.17$} \\     
         \hline            
        $I=10I_0$ & $3\omega_{mw}$ & $1.028$ &  $1.837$ & $0$ \\ 
         $(a_0=0.92)$& $5\omega_{mw}$ & $0.0765$ & $0.069$ & $0.464$ \\
         & $7\omega_{mw}$ & $0$ & \multicolumn{2}{|c|}{$0.21$} \\ 
         \hline
        \end{tabular}
        \caption{Conversion efficiency of reflected and transmitted harmonics of L and R polarisation has been shown for ($B_0=0.355T$)}
        \label{tab:intensity_transmittance}
    \end{table}
   
\section{Effect of Intensity of HPM pulse on harmonic generation}
   We have so far  explored the weakly nonlinear regime by  choosing  microwave intensity of about $I_0=5.857\times 10^{11}W/m^2$ for which relativistic factor corresponds to $a_0=0.29$. 
   Here, we have demonstrated the effect of microwave intensity on the generation of harmonics by making the intensity relativistic. Thus the regime of    ($a_0\approx  1 $) is explored.   Three different intensities $I=I_0, 5I_0, 10I_0$ of the microwave while keeping the other parameters fixed as chosen in case B have been chosen. For these intensities, figure (\ref{fig:Intensity_variation}) shows the profile of the transverse Electric field ($E_y^{mw}$) at $t= 15 ns$ inside the plasma for the corresponding three cases. By comparison, it could be understood that transmission of higher harmonic increases in comparison to the fundamental R-mode inside the magnetized plasma as the intensity is chosen to be relativistic. Time FFTs of these three cases have been shown in figure (\ref{fig:Intensity_fft}) at $x=2$ meters inside the plasma for a time window of $t=4ns$ to $20 ns$. The spectra clearly indicate that the efficiency of higher harmonics grows inside the plasma, and even $7^{th}$ harmonic shows a significantly high intensity. Figure (\ref{tab:intensity_transmittance}) shows the reflectivity and transmittance of higher harmonics in the vacuum and plasma, respectively, for the three different intensities. Reflected EM wave follows the linear dispersion relation ($\omega=kc$) and travels with the speed of light, making all the higher harmonics spatially superposed in the reflected region. While in the transmitted region in RL-mode configuration, RCP and LCP waves follow different dispersion relations and travel with different groups and phase velocities in the medium. Thus, both polarized waves get spatially separated. Since the polarised harmonics of R and L-wave have, We have calculated the efficiency of each polarized harmonic by the ratio of peaks of corresponding space FFT of each harmonic by the peak of incident incoming microwave in a vacuum. The above analysis in the table (\ref{tab:intensity_transmittance})  revealed that the efficiency of $3^{rd}$ harmonic of L-wave is finite, while for $3^{rd}$ harmonic of R-wave, it is zero. Similarly, the efficiency of $5^{th}$ harmonic of the R-wave gets enhanced in comparison to the L-wave component. While for the $7^{th}$ harmonic dispersion relation of R and L-mode almost becomes equivalent because of that it travels with linear polarisation. This illustrates that the alternate polarized harmonic got excited in the plasma medium. Further, in the second case, by increasing the intensity $5$ times, we observe that the efficiency of $3^{rd}$ harmonic of L-wave increases by $3.7$ times, while $5^{th}$ harmonic of R-wave increases $12.2$ times and $7^{th}$ harmonic grows almost by $80$ times. This elucidates that the efficiency of harmonics grows proportionally with the order of harmonics. This behavior agrees with our theoretical analysis in Appendix {\ref{currentdens}}. 

\section{Summary}\label{sec:summary}
 In this paper, the Interaction of High Power microwave with magnetized plasma has been comprehensively studied with one dimension PIC simulation. We have described the new results observed on the excitation of higher harmonic in RL mode configuration in case of an overdense plasma ($\omega_{mw}<\omega_{pe}$). It has been demonstrated that when the frequency of incident linearly polarized EM wave fall under particular regions of dispersion curve of R $\&$ L modes $({\tilde k||B_0})$, efficient higher odd harmonics of circularly polarized wave can be excited in the bulk plasma which travels with different group velocities. It has been demonstrated by choosing an appropriate external magnetic field in comparison to the incident microwave frequency. Harmonics of either RCP wave, LCP wave, or both kinds of polarized wave could be excited in the bulk plasma.
Further, for a circularly polarized EM wave, it has been established that higher harmonics cannot be generated in the bulk plasma and vacuum. The efficiency of harmonics could be further increased by choosing an external magnetic field such that it is closer to electron cyclotron resonance ($\omega_{mw}=\omega_{ce}$). By increasing the intensity of incoming electromagnetic waves, the conversion efficiency of higher harmonics grows in proportion to the order of the high harmonic. An earlier study by \cite{maity2021harmonic} revealed that propagating second harmonics could be excited in X and O mode configuration. However, in RL-mode configuration, symmetry restricts the generation of even harmonics at the surface. This leads to a couple the Electromagnetic energy transfer directly to the odd higher harmonics. The total maximum conversion efficiency of about 2.865 $\%$ has been detected for third harmonic radiation in the bulk plasma and vacuum for  $a_0=0.92$ and normalized external magnetic field of $B_0= 0.78$. It is also crucial that the efficiency of alternate polarization of harmonics grows in bulk plasma as the intensity of incident radiation increases. This could be further improved by optimizing intensity and external magnetic field parameters for chosen frequency of the electromagnetic wave. 

\section*{Acknowledgements}
 \indent The Authors would like to thank EPOCH consortium for providing open access to EPOCH 4.17.16 framework. AD would like to acknowledge her  J.C. bose fellowship grant of AD(JCB/2017/000055) as well as CRG/2018/000624, grant of Department of Science and Technology (DST) Government of India. The authors would like to thank IIT Delhi HPC facility for computational resources. T.D. would also wishes to thank Council for Scientific and Industrial Research (Grant no- 09/086/(1489)/2021-EMR-I) for funding the research. 
 
\section*{Conflict of Interest}
 \noindent Authors report no conflict of interest
 
\appendix

\section{Surface current oscillation in RL mode configuration}\label{appA}
  Let us now examine, effect of increasing the intensity of incoming microwave by keeping the external magnetic field to be fixed at $B_0=0.355T$. We have derived expressions for the surface current generated by electrons at the surface of plasma in RL-mode configuration. When the high power microwave incidents on overdense plasma electrons respond to the oscillating Electromagnetic field which in turn excite the higher harmonic for the particular range of $B_0$. 
 In the RL-mode configuration of incident microwaves, We have chosen external magnetic field and EM field in general form as, 
\begin{equation}
  {B_0}=B_0\hat{x}; \quad \tilde{E}_{mw}=E_{mw0}(x)(\hat{y}+\alpha \hat{z})exp(-i\omega_{mw}t)
  \label{Elec_field}
\end{equation}
Here $\alpha=0$ for linearly polarized wave and $\alpha=\pm \iota$ for LCP and RCP Wave.\\
\begin{equation}
  \nabla\times\tilde{E}=-\frac{\partial \tilde{B}}{\partial t}
  \label{Mag_field}
\end{equation}
Now since the variation of oscillating electric field is along x direction. So, from maxwell's third equation we have,
\begin{equation}
{\tilde{B}_{mw}}=-\frac{i}{\omega}\frac{\partial \tilde{E}_{mw0}}{\partial x}(\hat{z}-\alpha \hat{y})
\end{equation}
Now in vacuum x$\rightarrow$ ct, so we can write,
\begin{equation}
 {\tilde{B}_{mw}}=\frac{\tilde{E}_{mw0}}{c}(\hat{z}-\alpha \hat{y})exp(-i\omega_{mw}t)
\end{equation}
From the Lorentz force equation for electron in presence of EM field 
\begin{equation}
\frac{\partial\tilde{v}}{\partial t}=-\frac{e}{m}[\tilde{{E}}_{mw}+\tilde{{v}}\times ({B_0}+\tilde{{B}}_{mw})]
\end{equation}
Now first-order terms for velocity in (A5) we obtain

\begin{equation}
 \frac{\partial\tilde{v_x}^{(1)}}{\partial t}=0; \quad \frac{\partial\tilde{v_y}^{(1)}}{\partial t}=-\frac{e}{m}[E_{mwy}+\tilde{v_{z}}^{(1)}B_0]; \quad \frac{\partial\tilde{v_z}^{(1)}}{\partial t}=-\frac{e}{m}[E_{mwy}-\tilde{v_{y}}^{(1)}B_0]
\end{equation}
So in cold plasma since at t=0; $v_x=0$, So later also $v_x(t)=0$ thus we get,
\begin{equation}
 \tilde{v_x}^{(1)}=0; \quad \tilde{v_y}^{(1)}=-\frac{ie}{m\omega_{mw}}[E_{mwy}+\tilde{v_{z}}^{(1)}B_0]; \quad \tilde{v_z}^{(1)}=\frac{ie}{m\omega_{mw}}[E_{mwy}-\tilde{v_{y}}^{(1)}B_0]
\end{equation}
So by solving this coupled equation in $v_y^{(1)}$ and $v_z^{(1)}$ we get,
\begin{equation}
\tilde{v_y}^{(1)}=\frac{e}{m}\frac{(-i\omega_{mw}E_{mwy}-\omega_cE_{mwz})}{(\omega_{mw}^2-\omega_c^2)}; \quad \tilde{v_z}^{(1)}=\frac{e}{m}\frac{(-i\omega_{mw}E_{mwz}+\omega_cE_{mwy})}{(\omega_{mw}^2-\omega_c^2)}
\end{equation}
In terms of $\alpha$ we write,
\begin{equation}
\tilde{v_y}^{(1)}=\frac{eE_{mw0}}{m}\frac{(-i\omega_{mw} -\alpha \omega_c)}{(\omega_{mw}^2-\omega_c^2)}; \quad \tilde{v_z}^{(1)}=\frac{eE_{mw0}}{m}\frac{(-i\alpha\omega_{mw}+\omega_c)}{(\omega_{mw}^2-\omega_c^2)}
 \label{eq:velocity1st}
\end{equation}
In second order of velocity,
\begin{equation}
   \frac{\partial \tilde{v_x}^{(2)}}{\partial t}=-\frac{e}{m}[\tilde{v_y}^{(1)}B_{mwz}-\tilde{v_z}^{(1)}B_{mwy}] 
   \label{eq:vel_harmonic}
\end{equation} 

So substituting value for $\tilde{v_y}^{(1)}$ and $\tilde{v_z}^{(1)}$ from (\ref{eq:velocity1st}) we get,
\begin{equation}
    \tilde{v_x}^{(2)}=-\frac{e^2}{2m^2c}\frac{E_{mw0}^2}{(\omega_{mw}^2-\omega_c^2)}(1+\alpha^2)exp(-2i\omega_{mw}t)
    \label{velocity_2nd}
\end{equation}
 Thus Surface current along x-direction could be expressed as, 
 \begin{equation}
    \tilde{J_{ex}}^{(2)}=-\frac{n_e e^3}{2m^2c}\frac{E_{mw0}^2}{(\omega_{mw}^2-\omega_c^2)}(1+\alpha^2)exp(-2i\omega_{mw}t)
    \label{current_dens_2nd}
\end{equation}

Now for $\hat y$ and $\hat z$ component can be expressed as,
\begin{equation}
     \frac{\partial \tilde{v_y}^{(2)}}{\partial t}=-\frac{e}{m}[\tilde{v_z}^{(2)}B_{0}-\tilde{v_x}^{(1)}B_{mwz}]; \quad  \frac{\partial \tilde{v_z}^{(2)}}{\partial t}=-\frac{e}{m}[-\tilde{v_y}^{(2)}B_{0}+\tilde{v_x}^{(1)}B_{mwy}] 
\end{equation}
Since $\tilde{v_x}^{(1)}$=0; So this second order term will just give gyration in yz-plane. \begin{equation}
    \tilde{v_y}^{(2)}=\tilde{v_{y0}}^{(2)}exp(-i\omega_ct); \quad \tilde{v_z}^{(2)}=\tilde{v_{z0}}^{(2)}exp(-i\omega_ct);
\end{equation}
Now in 3rd order of velocity we write, 
\begin{equation}
     \frac{\partial \tilde{v_y}^{(3)}}{\partial t}=-\frac{e}{m}[\tilde{v_z}^{(3)}B_{0}-\tilde{v_x}^{(2)}B_{mwz}]; \quad  \frac{\partial \tilde{v_z}^{(2)}}{\partial t}=-\frac{e}{m}[-\tilde{v_y}^{(3)}B_{0}+\tilde{v_x}^{(2)}B_{mwy}] 
\end{equation}
Substituting for $v_x^{(2)}$ from (\ref{velocity_2nd}) we get coupled differential equations,
\begin{equation}
    \frac{\partial^2\tilde{v_y}^{(3)}}{\partial t^2}+\omega_c^2\tilde{v_y}^{(3)}=\frac{e}{m}\left[ \frac{\partial}{\partial t}(\tilde{v_x}^{(2)}B_{mwz})+\omega_c\tilde{v_x}^{(2)}B_{mwy} \right]
\end{equation}
\begin{equation}
    \frac{\partial^2\tilde{v_z}^{(3)}}{\partial t^2}+\omega_c^2\tilde{v_z}^{(3)}=\frac{e}{m}\left[ -\frac{\partial}{\partial t}(\tilde{v_x}^{(2)}B_{mwy})+\omega_c\tilde{v_x}^{(2)}B_{mwz} \right]
\end{equation}
We solve equations (A15,16) which can be expressed as,
\begin{equation}
    \tilde{v_y}^{(3)}=F(3i\omega_{mw}+\alpha \omega_c)exp(-i3\omega_{mw}t); \quad  \tilde{v_z}^{(3)}=F(3i\omega_{mw}\alpha- \omega_c)exp(-i3\omega_{mw}t)
\end{equation}
Here , 
\begin{equation}
    F=\frac{1}{2}\left(\frac{e}{m}\right)^3\frac{E_{mw0}^3(1+\alpha^2)}{c^2(\omega_{mw}^2-\omega_c^2)(\omega_c^2-9\omega_{mw}^2)}
\end{equation}
Now for circular polarisation $\alpha =\pm i $ we get from equation (A 18), F=0. So that means $\tilde{v_y}^{(3)}$=$\tilde{v_z}^{(3)}$=0. So that's why we haven't observed harmonics in our simulations for right and left circular polarisation. \\
While in case of linear polarisation $\alpha =0$  we get,
\begin{equation}
    \tilde{v_y}^{(3)}=F_1exp(-i3\omega_{mw}t); \quad  \tilde{v_z}^{(3)}=F_2exp(-i3\omega_{mw}t)
\end{equation}\label{velocity3rd}
Here, 
\begin{equation}
    F_1=\frac{1}{2}\left(\frac{e}{m}\right)^3\frac{3i\omega_{mw}E_{mw0}^3}{c^2(\omega_{mw}^2-\omega_c^2)(\omega_c^2-9\omega_{mw}^2)}\quad  F_2=-\frac{1}{2}\left(\frac{e}{m}\right)^3\frac{\omega_cE_{mw0}^3}{c^2(\omega_{mw}^2-\omega_c^2)(\omega_c^2-9\omega_{mw}^2)}
    \label{currentcoeff}
\end{equation}
So the current density in yz-plane could be expressed as,
\begin{equation}
    \tilde{J_{ey}}^{(3)}=-n_eeF_1exp(-i3\omega_{mw}t); \quad  \tilde{J_{ez}}^{(3)}=-n_eeF_2exp(-i3\omega_{mw}t)
    \label{currentdens}
\end{equation}
Equation (\ref{currentdens}) represents current density at the surface of plasma which oscillates with frequency $3\omega_{mw}$, when a linearly polarized light incidents on surface of plasma. \\

\section*{References}
\bibliographystyle{unsrt}
\bibliography{Harmonic.bib}

\end{document}